\documentclass[aps,apl,showpacs,twocolumn,superscriptaddress]{revtex4}
\usepackage{bm,color}
\usepackage{graphicx}
\usepackage{amsmath}
\pdfoutput=1

\begin{document}
\title {Controlled creation of nanometric skyrmions using external magnetic fields}
\author{Masahito Mochizuki}
\affiliation{Department of Applied Physics, Waseda University, Okubo, Shinjuku-ku, Tokyo 169-8555, Japan}
\affiliation{Department of Physics and Mathematics, Aoyama Gakuin University, Sagamihara, Kanagawa 229-8558, Japan}
\affiliation{PRESTO, Japan Science and Technology Agency, Kawaguchi, Saitama 332-0012, Japan}
\begin{abstract}
To exploit nanometric magnetic skyrmions as information carriers in high-density storage devices, a method is needed that creates intended number of skyrmions at specified places in the device preferably at a low energy cost. We theoretically propose that using a system with a fabricated hole or notch, the controlled creation of individual skyrmions can be achieved even when using an external magnetic field applied to the entire specimen. The fabricated defect turns out to work like a catalyst to reduce the energy barrier for the skyrmion creation.
\end{abstract}
\maketitle

\begin{figure}
\includegraphics[width=1.0\columnwidth]{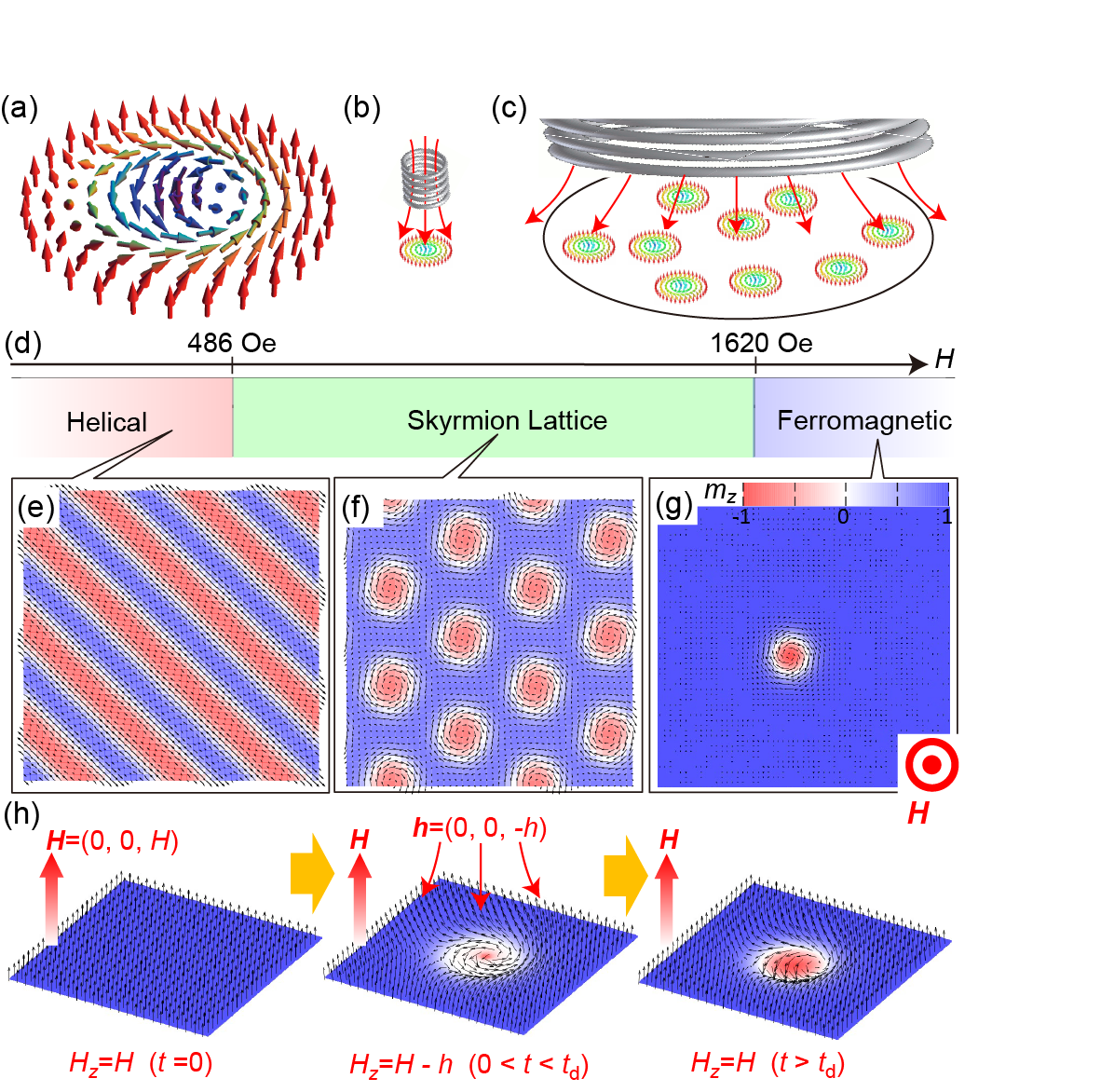}
\caption{(color online). (a) Swirling magnetic structure of a skyrmion. (b) Ideal but unrealistic way to write a skyrmion using a squeezed magnetic field. (c) Actual situation of skyrmion creation by the simple application of a magnetic field with spot size much larger than the size of the skyrmion. (d) Theoretical phase diagram of the spin model~(\ref{eqn:model}) at $T=0$ as a function of magnetic field applied normal to the plane. (e)--(g) Magnetic structures of (e) helical phase, (f) skyrmion-lattice phase, and (g) field-polarized ferromagnetic phase with a skyrmion as a defect. (h) Schematic of skyrmion creation in a magnetic-field application.}
\label{Fig01}
\end{figure}

Magnetic skyrmions, nanometric swirling spin textures carrying a quantized topological number, in chiral-lattice ferromagnets are currently attracting enormous research interest~\cite{Rossler06,Fert13,Nagaosa13,Mochizuki15,Seki15}. The absence of spatial inversion symmetry in chiral crystals activates the Dzyaloshinskii--Moriya (DM) interaction, which favors a rotating alignment of the spin magnetizations. This interaction strongly competes with the ferromagnetic exchange interaction $J$ favoring a parallel alignment of magnetizations, and stabilizes the vortex-like skyrmion spin textures under an external magnetic field~\cite{Bogdanov89,Bogdanov94}. The magnetizations of a skyrmion are parallel to the magnetic field at its periphery, whereas they are antiparallel at its center to maximize energy gains from the Zeeman interaction [see Fig.~\ref{Fig01}(a)].

Skyrmions have numerous advantages for applications in data storage devices~\cite{Fert13,Nagaosa13}. First, they are nanometrically small in size (typically 3--100~nm), enabling high densities of information. Second, they are highly mobile, which facilitates their manipulation, thereby reducing the energy expended. Skyrmions can be driven, for example, by spin-polarized electric currents via the spin-transfer torque mechanism. The typical threshold electric current density $j_{\rm c}$ to drive them turns out to be 10$^5$--10$^6$~A/m$^2$~\cite{Jonietz10,YuXZ12,Iwasaki13a,Schulz12,Everschor11,ZangJ11,Reichhardt15}, which is five or six orders of magnitude smaller than $j_{\rm c}$ to drive other magnetic textures. Third, skyrmions are topologically protected and thus stable. They cannot be created and annihilated by continuous variation of the spatial alignment of magnetizations, but a discontinuous 180$^{\circ}$ flop of the local magnetization is necessary to write or delete them. This local magnetization flop expends a large energy cost of order $J$, which makes the written skyrmions robust against agitations.

To exploit magnetic skyrmions as information carriers in data storage devices, efficient methods to write, delete, read, and drive skyrmion bits need to be established. Regarding the writing of skyrmions, several techniques using electric currents~\cite{Iwasaki13b,Sampaio13,ZhouY14,JiangW15}, magnetic fields~\cite{Koshibae15,Buttner15,SWoo16}, electric fields~\cite{Mochizuki15a,Mochizuki15b,Okamura16}, spin currents~\cite{Romming13}, light~\cite{Finazzi13,Koshibae14}, heat~\cite{Oike16}, spin waves~\cite{LiuY15}, and pressure~\cite{Nii15} have been proposed from theory and experiment. Among these proposals, a method based on the application of a magnetic field holds promise as a practical means because the writing technique using a magnetic field has a long research history and accumulated knowledge in the development of hard-disk drives. 

However, creating an intended number of nanometric skyrmions at specific places using a magnetic field is not straightforward. This is because a magnetic field cannot be focused into an area as small as the skyrmion. Specifically, we ideally desire tight focusing as depicted in Fig.~\ref{Fig01}(b) but in actuality the field spreads out [Fig.~\ref{Fig01}(c)]. In addition, we need to establish a technique for writing a topological skyrmion texture without the large energy expense incurred with local magnetization flops.

In this Letter, we investigate a method to create skyrmions on a thin film of a chiral-lattice ferromagnet with an applied magnetic field based on numerical simulations using the Landau--Lifshitz--Gilbert (LLG) equation. We find that even when a magnetic field is applied to an entire thin-film specimen, it is possible to create individual skyrmions one by one within an intended nanoscopic area by fabricating nanoscopically small defects such as hole and notch on the sample. The threshold magnetic field turns out to be significantly lower compared with that for defect-free specimens. Mechanisms for skyrmion creation and the reduction of the threshold field are presented based on the simulation results. Our proposal, the controlled creation of nanoscopic spin textures via global field application, may provide a breakthrough in the research on high-density storage devices with ultra-small magnetic textures as information bits.

Magnetism in a thin-film specimen of chiral-lattice ferromagnet is described using the classical Heisenberg model on a square lattice with ferromagnetic exchange interaction, the DM interaction, and the Zeeman interaction associated with an external magnetic field $\bm H$=(0, 0, $H$) normal to the plane~\cite{Bak80,YiSD09}. The Hamiltonian is given as
\begin{eqnarray}
\mathcal{H}_0&=&
-J \sum_{<i,j>} \bm m_i \cdot \bm m_j
-D \sum_{i,\hat{\bm \gamma}} 
\bm m_i \times \bm m_{i+\hat{\bm \gamma}} \cdot \hat{\bm \gamma} 
\nonumber \\
&-&g\mu_{\rm B}\mu_0 \bm H \cdot \sum_i \bm m_i,
\label{eqn:model}
\end{eqnarray}
where $g$=2, and $\hat{\bm \gamma}$ denotes collectively the unit vectors $\hat{\bm x}$, $\hat{\bm y}$, and $\hat{\bm z}$ for the square lattice. The length of the magnetization vectors $\bm m_i$ is taken to be unity. We take $J$=3~meV and $D/J$=0.09 so as to reproduce the experimentally observed $T_{\rm c}$ and the typical periodicity ($\sim$ 50~nm) for the skyrmion lattice in the skyrmion material Cu$_2$OSeO$_3$~\cite{Seki12a,Adams12}. The magnetic field $\bm H$ is applied uniformly and steadily over the entire system.

The $H$-dependent phase diagram for this spin model [Fig.~\ref{Fig01}(d)] shows three phases, specifically, helical phase [Fig.~\ref{Fig01}(e)], skyrmion-lattice phase [Fig.~\ref{Fig01}(f)], and field-polarized ferromagnetic phase. These phases reproduce experimental results at low temperature. Importantly, skyrmions appear not only as the crystalized form in the skyrmion-lattice phase~\cite{Muhlbauer09,YuXZ10,Tonomura12} but also as individual defects in the ferromagnetic state~\cite{YuXZ10}; see Fig.~\ref{Fig01}(g). Isolated skyrmions in the ferromagnetic background are potentially useful as information carriers.

The LLG equation is given by
\begin{equation}
\frac{d\bm m_i}{dt}=-\bm m_i \times \bm H^{\rm eff}_i 
+\frac{\alpha_{\rm G}}{m} \bm m_i \times \frac{d\bm m_i}{dt},
\label{eq:LLGEQ}
\end{equation} 
where $\alpha_{\rm G}$(=0.04) is the Gilbert-damping coefficient. The effective field $\bm H_i^{\rm eff}$ is calculated from the Hamiltonian $\mathcal{H}$=$\mathcal{H}_0$+$\mathcal{H}^{\prime}(t)$ as
$\bm H^{\rm eff}_i = -\partial \mathcal{H} / \partial \bm m_i$.
The first term $\mathcal{H}_0$ is the model Hamiltonian given by Eq.~(\ref{eqn:model}). The second term $\mathcal{H}^{\prime}(t)$ represents the Zeeman interaction associated with a temporal magnetic field $\bm h(t)$=(0, 0, $-h$) applied in the direction opposite to the steady field $\bm H$ to create (write) a skyrmion. The term is given by,
\begin{eqnarray}
\mathcal{H}^{\prime}(t)=-g\mu_{\rm B} \mu_0 \sum_i \bm h(t) \cdot \bm m_i.
\label{eqn:Hpulse}
\end{eqnarray}
A process of the skyrmion creation is schematically shown in Fig.~\ref{Fig01}(h). The writing field $\bm h(t)$ is applied to a thin-film system with ferromagnetically ordered magnetizations polarized by the steady field $\bm H$. The writing field locally reverses the magnetization to form a skyrmion core, which after switching off the writing field grows into an isolated skyrmion via relaxation of the magnetization alignment. We simulated the spatio-temporal dynamics of these magnetizations $\bm m_i$ over this entire process. The simulations were performed using a system of $N$=600$\times$150 sites set with an open boundary condition. The strength of the steady field $\bm H$ is fixed at $H$=1622 Oe to ensure the system is ferromagnetic on the verge of the phase boundary to the skyrmion-lattice phase. We examine several cases of different field-applied areas and various fabricated structures.

A process of the skyrmion creation is schematically shown in Fig.~\ref{Fig01}(h). The writing field $\bm h(t)$ is applied to a thin-film system with ferromagnetically ordered magnetizations polarized by the steady field $\bm H$, which reverses the magnetization locally to form a skyrmion core. This seed of skyrmion grows to an isolated skyrmion via relaxation of the magnetization alignment after switching off the writing field $\bm h(t)$. We simulate spatio-temporal dynamics of magnetizations $\bm m_i$ in the skyrmion-creation process after applying $\bm h(t)$ from $t$=0. The simulations are performed using a system of $N$=600$\times$150 sites with the open boundary condition. The strength of the steady field $\bm H$ is fixed at $H$=1622 Oe to make the system ferromagnetic on the verge of the phase boundary to the skyrmion-lattice phase. We examine several cases of different field-applied areas and various fabricated structures.

\begin{figure}
\includegraphics[width=1.0\columnwidth]{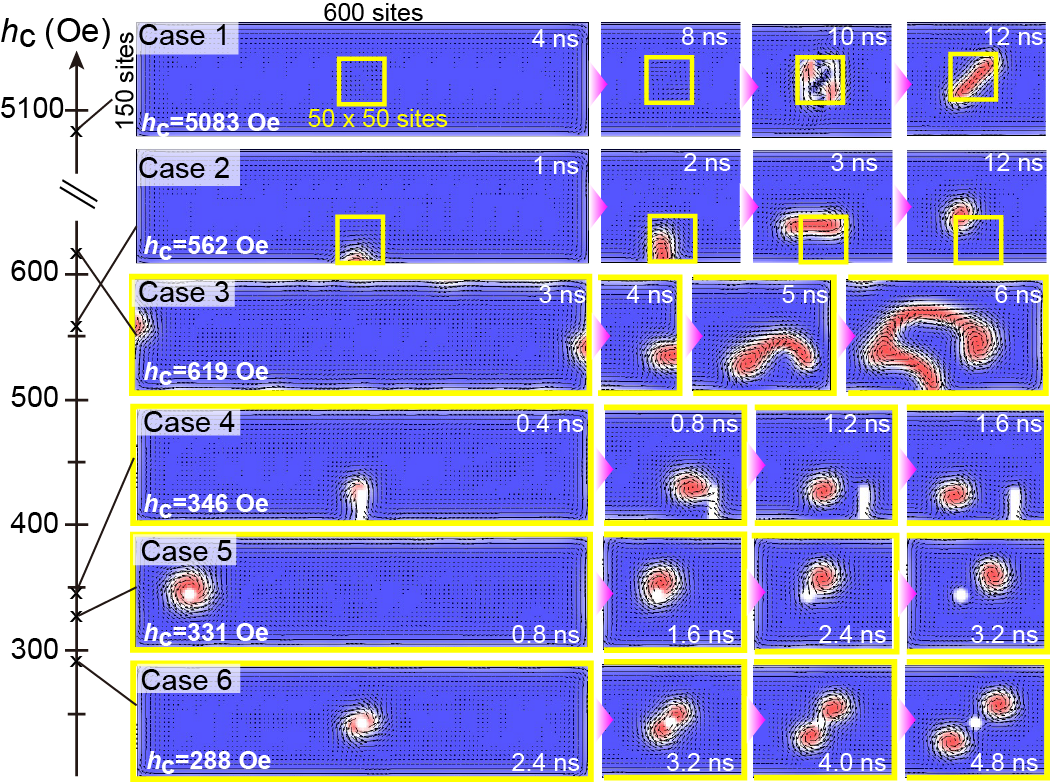}
\caption{(color online). Simulated threshold magnetic fields $h_{\rm c}$ for different settings in skyrmion creation. Areas where the writing field $\bm h$ is applied are framed using (yellow) squares or rectangles. Time profiles of the creation process are also displayed where the relevant areas are magnified. For Cases 4--6, the fabricated hole and notch structures in the thin film enable a one-by-one controlled creation of skyrmions at low threshold fields even with the global application of the writing field $\bm h$. The height and width of the rectangular notch in Case 4 are 45 sites and 15 sites, respectively. For Cases 5 and 6, the hole diameter is 18 sites.}
\label{Fig02}
\end{figure}
In Fig.~\ref{Fig02}, the simulated threshold fields $h_c$ for various settings are summarized together with time-space profiles of the skyrmion creation processes. Here the field-applied areas are marked by (yellow) squares. When the writing field $\bm h$ is applied locally at the center of the system as in Case 1, a large field is needed to realize a 180$^{\circ}$ flop of the local magnetization with $h_c$ reaching nearly 5100 Oe. It was theoretically proposed that the required strength for $\bm h$ to inscribe a skyrmion can be reduced by applying the field at the edge of the sample, as in Case 2. This is because topological protection is relaxed because the magnetization distribution at the edge is discontinuous, and a gradual rotation of the magnetization towards a 180$^{\circ}$ reversal becomes possible. However, although this proposal is conceptually interesting, the method is unrealistic and impractical because it is technically difficult or physically impossible to squeeze the magnetic field $\bm h$ into a nanoscopic area.

Instead, the real situation is to apply the magnetic field $\bm h$ to an area much larger than the skyrmion size. If the field $\bm h$ is applied to the entire area of the sample, the magnetization rotations take place somewhere on the edges, but we find that the number of created skyrmions, the locations of their creation, and even the shape of the created magnetic textures are not controllable, as in Case 3. This indicates that the simple application of a magnetic field to the entire area does not lead to a controlled creation of skyrmions, in contrast to the local application to a nanoscopic area that is physically impossible. How then, through an application of a magnetic field, can we create skyrmions in a controlled manner?

This seemingly unsolvable problem is resolved by fabricating a nanoscopic defect on the thin-film sample. With a small rectangular notch on the sample [Fig.~\ref{Fig02}, Case 4], skyrmions are created one by one at the notch even if we apply a magnetic field $\bm h$ over the entire system. Importantly, because the topological protection at the notch is relaxed, the threshold field strength is significantly reduced by one order of magnitude below that for Case 1. The time interval between successive skyrmion creations is governed by the strength of $\bm h$, and we can create the intended number of skyrmions by tuning the field strength and duration of the field application.

We can also expect this kind of controlled skyrmion creation even with a nanoscopic hole. When a magnetic field $\bm h$ is applied to a thin film with a circular hole (Case 5), skyrmions are again created at the hole. Interestingly, when the hole is located deep inside the sample (Case 6), a pair of skyrmions is created simultaneously.

\begin{figure}
\includegraphics[width=1.0\columnwidth]{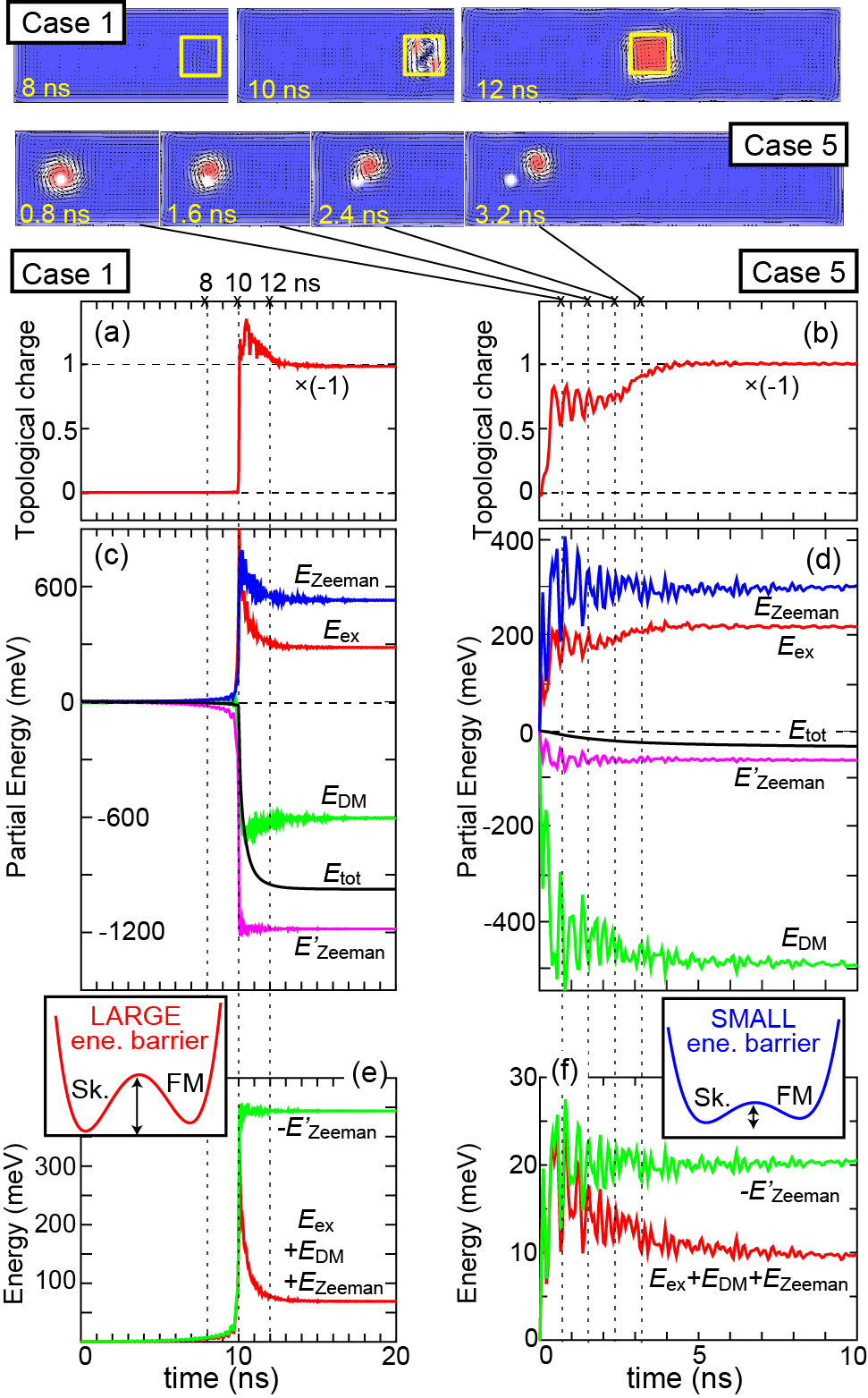}
\caption{(color online). (a), (b) Simulated time evolutions of topological charge for Case 1 (a) and Case 5 (b). (c)-(f) Simulated time profiles of partial energies for Case 1 [(c), (e)] and Case 5 [(d), (e)].}
\label{Fig03}
\end{figure}
The observed reduction of threshold field for Cases 3--5 is evident by comparing the results with those of Case 1. Figure~\ref{Fig03}(a) and (b) shows simulated time profiles of the topological charge for Cases 1 and 5, respectively. When the magnetic field $\bm h$ is locally applied to the system without a notch or hole (Case 1), a skyrmion is created suddenly accompanied by an abrupt flop in the local magnetization. Although we start applying the field $\bm h$ from $t$=0, almost nothing happens until $t$$\sim$10~ns. In contrast, when the field is applied completely over the entire system plus hole, the topological charge gradually increases from $t$=0, indicating a continuous rotation of magnetizations towards formation of the skyrmion core.

Figure~\ref{Fig03}(c) and (d) shows the simulated time evolution of partial energies for Cases 1 and 5, respectively. We find that the energy of the ferromagnetic-exchange interaction ($E_{\rm ex}$) and that of the Zeeman interaction associated with the steady field $\bm H$ ($E_{\rm Zeeman}$) increase for both cases as the uniformly aligned ferromagnetic configuration has been broken, whereas the energy of the DM interaction ($E_{\rm DM}$) and that of the Zeeman interaction associated with the writing field $\bm h$ ($E^{\prime}_{\rm Zeeman}$) decrease as the magnetization is canted towards the $\bm h$-direction. We find that these changes in the energy profiles abruptly occur in Case 1, whereas in Case 2 they gradually occur beginning after the application of $\bm h$. Importantly, the energy scale is much larger for Case 1 than for Case 2.

Figure~\ref{Fig03}(e) and (f) shows other results of time profiles of partial energies for Cases 1 and 5, respectively. The energy associated with the stationary part of the Hamiltonian $\mathcal{H}_0$, i.e., $E_{\rm ex}+E_{\rm DM}+E_{\rm Zeeman}$ shows critical enhancement in Case 1 at the skyrmion creation point, indicating the presence of a large energy barrier between states with and without a skyrmion. In this case, to create the skyrmion, we need to apply a much stronger writing field $\bm h$ so as the Zeeman-energy gain ($E^{\prime}_{\rm Zeeman}$) exceeds this high-energy barrier. In contrast, there appears no such energy enhancement in the skyrmion creation process for Case 5, indicating that the energy barrier, which must be overcome to create a skyrmion core, is very low. Consequently, the required strength of the writing field $\bm h$ is much weaker than that for Case 1. The fabricated defect works like a catalyst to reduce the energy barrier between the two states.

\begin{figure}[h]
\includegraphics[width=1.0\columnwidth]{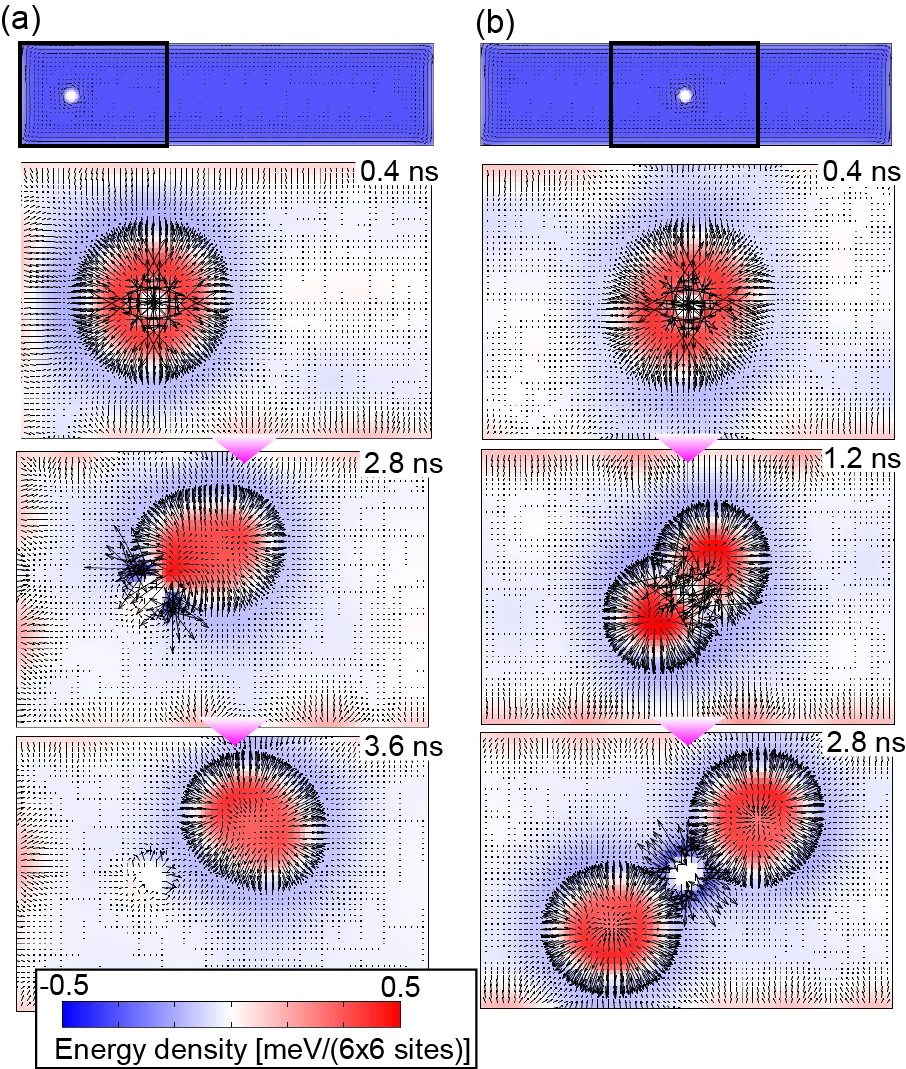}
\caption{(color online). Snapshots of the simulated spatial distribution of energy (color scaled) and the potential gradient (arrows) for Case 5 (a) and Case 6 (b), respectively. The relevant areas framed by rectangles of 225 $\times$ 150 sites are magnified. The strength of the writing field is $h$=331 Oe for both cases.}
\label{Fig04}
\end{figure}
Finally, we describe a mechanism for desorbing the generated skyrmion seed from the fabricated nanoscopic defect. Figure~\ref{Fig04}(a) displays snapshots of the simulated spatial distributions of the potential gradient, that is, the forces acting on each portion of the magnetic texture for Case 5. Because of repulsive forces from the sample edge, the distribution of the forces is anisotropic, which works to desorb the generated skyrmion seed from the hole and results in the formation of an individual skyrmion. Conceivably, if the hole is located at the center of the sample, the skyrmion core might not be desorbed from the hole because of an isotropic distribution of the potential gradient. However, as seen in Fig.~\ref{Fig04}(b), even when the hole is located at the center of the sample, the growing skyrmion core splits into two, and they are desorbed from the hole through their mutually repulsive force.

To summarize, we have proposed that controlled one-by-one creation of nanometric topological skyrmion spin textures with low energy cost is possible using a system with a fabricated nano-scale defect even with a global application of the magnetic field. The physical mechanism is ascribed to the relaxation of the topological protection because of the discontinuity in the spatial distribution of magnetization around the defect, which enables a gradual magnetization rotation instead of a sudden 180$^{\circ}$ flop to create a skyrmion core. To further increase the information density in magnetic memory devices, magnetic structures carrying information must be nanoscopically small, but it is technically difficult or physically impossible to achieve nanoscale-sized spots of magnetic field, light, and heat. Our proposal for a controlled creation of nanoscopic magnetic textures through the global application of an external field may provide the breakthrough needed to increase information densities of magnetic storage devices. The method can be applicable not only to the Bloch-type skyrmions in chiral magnets treated in the present study but also to the N\'eel-type skyrmions in interfacial systems~\cite{Fert13,Sampaio13,Tomasello14,Soumya16} and polar magnets~\cite{Kezsmarki15}. In addition to the data storage devices, a number of possible applications of magnetic skyrmions have been proposed to date~\cite{Finocchio16} such as logic~\cite{ZhangX15}, microwave~\cite{Mochizuki13,Okamura13,Finocchio15} and nano-oscillator devices~\cite{LiuRH15,ZhouY15b,Carpentieri15,ZhangS15}. The proposed method may be useful for these applications as well.

The author thanks A. Takeuchi and H. Nakayasu for discussions. This research was supported by JSPS KAKENHI (Grant Nos. 25870169, 25287088, and 17H02924), Waseda University Grant for Special Research Projects (Project No. 2017S-101), and JST PRESTO (Grant No. JPMJPR132A). 

\end{document}